\documentclass[useAMS,galley]{mn2e}
\usepackage{graphicx}
\usepackage{times,mathptm}
\usepackage{lineno}
\newcommand{\ltsima}{$\; \buildrel < \over \sim \;$}
\newcommand{\simlt}{\lower.5ex\hbox{\ltsima}}
\newcommand{\gtsima}{$\; \buildrel > \over \sim \;$}
\newcommand{\simgt}{\lower.5ex\hbox{\gtsima}}

\newcommand{\cgs}{ ${\rm erg~cm}^{-2}~{\rm s}^{-1}$} 
\newcommand{\lum}{\rm erg~s$^{-1}$}

\def\lesssim{\mathrel{\hbox{\rlap{\hbox{\lower4pt\hbox{$\sim$}}}\hbox{$<$}}}}
\def\gtrsim{\mathrel{\hbox{\rlap{\hbox{\lower4pt\hbox{$\sim$}}}\hbox{$>$}}}}

\def\arcsec{\hbox{$^{\prime\prime}$}}

\def\aox{$\alpha_{\rm ox}$}

\def\ab1450{$AB_{1450(1+z)}$}

\def\xray{\hbox{X-ray}}

\def\oiii{\hbox{[O\ {\sc iii}}]}
\def\ovii{\hbox{O\ {\sc vii}}}
\def\oviii{\hbox{O\ {\sc viii}}}

\def\civ{\hbox{C\ {\sc iv}}}

\newcommand\phn{\phantom{0}}%
\def\pg{PG~1543$+$489}

\def\msun{M$_{\odot}$}
\def\edd_ratio{$\log\ L_{\rm bol}/L_{\rm Edd}$}
\def\asca{{\it ASCA\/}}
\def\chandra{{\it Chandra\/}}

\def\heao1{{\it HEAO-1\/}}

\def\rosat{{\it ROSAT\/}}
\def\rxte{{\it RXTE\/}}
\def\sax{{\it BeppoSAX\/}}

\def\xmm{{XMM-{\it Newton\/}}}
\def\suzaku{{\it Suzaku\/}}

\def\aj{AJ}
\def\araa{ARA\&A}
\def\apj{ApJ}

\def\apjs{ApJS}

\def\aap{A\&A}

\def\mnras{MNRAS}

\def\an{Astron. Nachr.}
\def\nat{Nature}

\def\gca{Geochim.~Cosmochim.~Acta}
 
\title[THE STRANGE CASE OF PG~1543$+$489]
{On the peculiar properties of the narrow-line quasar PG~1543$+$489} 
\author[C. Vignali et al.]
{
Cristian Vignali,$^{1,2}$\thanks{E-mail: cristian.vignali@unibo.it (CV); 
piconcelli@mporzio.astro.it (EP); bianchi@fis.uniroma3.it (SB); miniutti@apc.univ-paris7.fr (GM).} 
Enrico Piconcelli$^{3}$\footnotemark[1],
Stefano Bianchi$^{4,5}$\footnotemark[1] and 
Giovanni Miniutti$^{6,7}$\footnotemark[1] \\ \\ 
$^{1}$ Dipartimento di Astronomia, Universit\`a degli Studi di Bologna, 
Via Ranzani 1, I--40127 Bologna, Italy \\
$^{2}$ INAF -- Osservatorio Astronomico di Bologna, Via Ranzani 1, 
I--40127 Bologna, Italy \\
$^{3}$ INAF -- Osservatorio Astronomico di Roma, Via Frascati 33, I--00040 Monteporzio Catone, 
Italy \\
$^{4}$ Dipartimento di Fisica, Universit\`a degli Studi di Roma Tre, Via della Vasca Navale 84, 
I--00146 Roma, Italy \\
$^{5}$ ESA -- European Space Astronomy Center, Apartado 50727, E--28080 Madrid, Spain \\
$^{6}$ Institute of Astronomy, Madingley Road, Cambridge CB3 0HA, UK \\
$^{7}$ Laboratoire APC, UMR 7164, 10 rue A. Domon et L. Duquet, 75205 Paris, France
}

\begin{document}

\date{Accepted 2008 May 7.  Received 2008 May 7; in original form 2008 March 21}
\pagerange{\pageref{firstpage}--\pageref{lastpage}} \pubyear{2006}

\maketitle

\label{firstpage}

\begin{abstract}
We present the analysis of four \xmm\ observations of the narrow-line quasar \pg\ at $z$=0.400 
carried out over a rest-frame time-scale of about three years. 
The \xray\ spectrum is characterized by a broad, relativistic iron K$_{\alpha}$ emission line 
and a steep photon index, which can be both explained by a ionized reflection model, 
where the source of \xray\ photons is presumably very close to the black hole. 
If this were the case, strong light-bending effects are expected, and actually they provide 
the most plausible explanation for the large equivalent width (EW=3.1$\pm{0.8}$~keV in the source rest frame) 
of the iron line. 
Although the light-bending model provides a good description of the \xray\ data of \pg, 
it is not possible to rule out an absorption model, where obscuring matter partially covers the \xray\ source. 
However, the apparent lack of variations in the properties of the absorber 
over the time-scale probed by our observations may indicate that this model is less likely. 
\end{abstract}


\section{Introduction}
\label{introduction}
The Palomar-Green (PG) quasars (Schmidt \& Green 1983) represent one of the best studied samples of 
Active Galactic Nuclei (AGN). Because of their relatively high \xray\ luminosities, 
over the last decade their properties have been widely investigated by almost 
all the \xray\ satellites, from \rosat\ (e.g., Laor et al. 1997) to \asca\ 
(e.g., George et al. 2000), \sax\ (e.g., Mineo et al. 2000) and, lastly, 
\xmm\ (Porquet et al. 2004; Piconcelli et al. 2005; Jim{\'e}nez-Bail{\'o}n et al. 2005; 
Brocksopp et al. 2006). In particular, the high-energy throughput of the \xmm\ EPIC instruments 
have recently allowed a detailed investigation of their spectral properties in the 
\hbox{$\approx$~0.3--10~keV} band (e.g., Piconcelli et al. 2005; 
Jim{\'e}nez-Bail{\'o}n et al. 2005), showing the ubiquitous presence of soft excesses and 
iron K$_{\alpha}$ emission lines, as well as, in half of the sample, 
of warm absorbers. 

Here we present the \xray\ analysis of the \xmm\ spectrum of \pg, a narrow-line quasar (NLQ) 
[the full width at half maximum (FWHM) of the H$_\beta$ line is 1630~km~s$^{-1}$ (Aoki, Kawaguchi \& Ohta 2005)] 
at a redshift of $z$=0.400. 
According to optical and ultra-violet mass scaling, this quasar is likely characterized by 
a (1--2.4)$\times10^{8}$~\msun\ black hole (D'Elia, Padovani \& Landt 2003; Vestergaard \& Peterson 2006); 
a even larger black hole mass was estimated by Brocksopp et al. (2006) using a 
spectral energy distribution (SED) fitting approach. 
Furthermore, from the measured $\lambda$$L_{5100\mbox{\scriptsize\AA}}$ ($\approx$~10$^{45.6}$~\lum; 
Aoki et al. 2005) and adopting the bolometric correction for broad-line (Type~1) quasars reported in 
Richards et al. (2006), we can estimate a remarkably high Eddington 
ratio [defined as L$_{\rm bol}$/L$_{\rm Edd}$, where L$_{\rm bol}$ is the bolometric luminosity 
and L$_{\rm Edd}$=(1.3--3.1)$\times10^{46}$~\lum\ is the Eddington luminosity] 
of $\approx$~1.3--3.7 for \pg\ (vs. $\approx$~2.3 of Baskin \& Laor 2005). 
This value, similar to that obtained by Aoki et al. (2005), 
appears significantly higher than the one estimated from the observed 2--10~keV luminosity 
(\hbox{$\approx1.1\times10^{44}$~\lum}; see $\S$\ref{longterm_flux}) 
using the Elvis et al. (1994) average SED of broad-line quasars, which is $\approx$~0.1--0.3. 
The difference in the estimates of the bolometric luminosity (from the $\lambda$$L_{5100\mbox{\scriptsize\AA}}$ 
and the 2--10~keV luminosity) can be partially overcome by assuming 
Eq.~21 in Marconi et al. (2004; see their section~3.2). 
In this case, the bolometric luminosity derived using the B$-$band luminosity is 
a factor of $\approx$~4.8 lower than the one obtained by Richards et al. (2006) 
from the $\lambda$$L_{5100\mbox{\scriptsize\AA}}$. 
%
%
We expect that in objects with large Eddington ratios the metallicity is considerably high 
(e.g., Shemmer et al. 2004; Netzer \& Trakhtenbrot 2007), as actually observed in \pg\ 
(Aoki et al. 2005). 

A peculiarity of \pg\ is the blueshift of the \oiii\ 5007\AA\ line (1150~km~s$^{-1}$ with 
respect to the systemic velocity of the galaxy) and the blue asymmetry of its profile 
(Aoki et al. 2005). The large \oiii\ blueshift of the 
so-called ``blue outliers''\footnote{According to the definition of Marziani et al. 2003b, 
a ``blue outlier'' is a source where 
$\Delta v_{r\ [OIII]} = v_{r}([OIII]\lambda5007) - v_{r}(H_\beta) \simlt -300$~km~s$^{-1}$, 
where $v_{r}$ is the radial velocity as measured on the Fe\ II--subtracted spectrum.}
(to which \pg\ belongs) has been theoretically interpreted as the result of an outflow 
whose receding part is obscured by an optically thick accretion disc (Zamanov et al. 2002) or 
by a scenario in which the narrow-line region clouds are entrained in a decelerating wind, 
possibly linked to the high Eddington ratio typical of the ``blue outliers'' (Komossa et al. 2008). 
The amount of blueshift does not appear to correlate with the Eddington ratio; however, there is 
a clear connection between being a ``blue outlier'' and the Eddington ratio itself (Aoki et al. 2005); 
this issue clearly needs further and extended investigations.  
Interestingly, \pg\ has the largest \oiii\ blueshift (relative to H$_\beta$) among the 280 
broad-line AGN in the sample of Marziani et al. (2003a); 
although outflow phenomena have been reported in other narrow-line Seyfert~1 galaxies (NLS1s) and NLQs, 
to our knowledge \pg\ shows the most extreme value of blueshift among AGN not classified 
as broad absorption-line (BAL) quasars. 
The remarkably strong asymmetric profile of the blueshifted \civ\ emission line (Baskin \& Laor 2005) 
provides a further indication that \pg\ is an intriguing source with a large Eddington ratio. 
%
\begin{table*}
\centering
\begin{minipage}{0.93\textwidth}
\caption{\xmm\ observation log.}
\label{obs_log}
\
\begin{tabular}{cccccccccc}
\hline
Quasar & RA        & DEC       & $z$ & $N_{\rm H}$$^{a}$ & Observation & Start Date & \multicolumn{3}{c}{Net Exposure Time$^{b}$ / Source Counts$^{c}$ / Extr. Radius$^{d}$} \\
Name   & (J2000.0) & (J2000.0) &     &                   & OBS$\_{\rm ID}$ & & pn & MOS1 & MOS2 \\
\hline
PG~1543$+$489 & 15 45 30.2 & 48 46 09.1 & 0.400 & 1.59 & 0153220401 & 2003 Feb 08 & {\phn}9.3/3010/27 & 11.6/860/24  & 11.8/910/24 \\
              &            &            &       &      & 0505050201 & 2007 Jun 09 & {\phn}6.6/3530/45 &  8.8/1030/30 &  8.8/950/30 \\
              &            &            &       &      & 0505050701 & 2007 Jun 15 & {\phn}6.7/3160/35 &  8.8/950/25  &  8.7/920/25 \\
              &            &            &       &      & 0505050301 & 2007 Jun 17 & 10.4/4730/25      & 13.6/1520/25 & 13.6/1480/25 \\
\hline
\end{tabular}
\end{minipage}
\begin{minipage}{0.93\textwidth}
$^{a}$ Neutral Galactic absorption column density in units of $10^{20}$~cm$^{-2}$ 
obtained from Dickey \& Lockman (1990). 
$^{b}$ The exposure times (ks) have been corrected for the high-background intervals 
(see $\S$\ref{data_red} for details), which still contaminate
significantly the data of OBS$\_{\rm ID}$=0505050301. 
$^{c}$ The source counts are reported in the \hbox{0.3--10~keV} band. 
$^{d}$ Source extraction radius (arcsec). 
\end{minipage}
\end{table*}

%

Previous observations in the soft and hard X-rays have shown that NLS1s and NLQs have typically steeper 
\xray\ spectra [in both the soft (e.g., Boller, Brandt \& Fink 1996) and hard \xray\ band (e.g., Brandt, Mathur \& Elvis 1997)]
than ``normal'' broad-line quasars, whose photon index 
is \hbox{$\Gamma\approx$~1.9--2.0} in the 2--10~keV band (e.g., Reeves \& Turner 2000; Page et al. 2003) and independent 
on the source redshift and luminosity (e.g., Piconcelli et al. 2003; Vignali et al. 2005; 
Shemmer et al. 2005; Page et al. 2005; but see also Dai et al. 2004 and Saez et al. 2008 for different results). 
%
In Type~1 AGN, a well established anti-correlation between the photon index 
and the H$_\beta$ FWHM exists both in the soft and hard \xray\ band (e.g., Laor et al. 1994, 1997; 
Brandt et al. 1997). 
Based on the reasoning that eventually led to the Kaspi et al. (2000) results, 
the FWHM of the H$_\beta$ line was suggested to be an accretion rate indicator in broad-line AGN 
(e.g., Boroson \& Green 1992; Brandt \& Boller 1998), i.e., 
objects with narrower H$_\beta$ lines 
are thought to accrete at a higher fraction of the Eddington limit. 
On the basis of this finding, Reeves \& Turner (2000) 
explained the steepening of the \xray\ spectrum as due to a larger 
Compton cooling of the hard \xray\ emitting corona (e.g., Pounds, Done \& Osborne 1995; 
see also Puchnarewicz et al. 1995). 

In this context, the \xray\ observations of \pg\ are meant to provide a powerful insight on 
the emission in the innermost regions of the quasar; the presence of a steep photon index, 
as indicated by the \asca\ observation (George et al. 2000), coupled with the narrow H$_\beta$ 
emission line, would imply a high accretion rate for this source (see the recent works by Shemmer 
et al. 2006, 2008 and references therein), likely related to the outflow phenomena. 

Hereafter we adopt \hbox{$H_{0}$=70~km~s$^{-1}$~Mpc$^{-1}$} in a $\Lambda$-cosmology 
with \hbox{$\Omega_{\rm M}$=0.3} and \hbox{$\Omega_{\Lambda}$=0.7} (Spergel et al. 2003).

\section{XMM-Newton observations of \pg}
\label{xmm_obs_log}
The \xmm\ data reported in this paper consist of four observations, the first of 
which (dated Feb. 2003) being retrieved from the archive and published 
by Matsumoto, Leighly \& Kawaguchi (2006) and Brocksopp et al. (2006). 
The remaining three observations were obtained in \xmm\ AO6 call for proposals, 
and the source was observed in June 2007. 
The observation log of all the \xmm\ observations of \pg\ is reported in Table~\ref{obs_log}. 

\subsection{EPIC data reduction}
\label{data_red}
The \xmm\ data were processed using standard {\sc sas v7.0.0} (Gabriel et al. 2004) 
and {\sc ftools} tasks. 
The event files were filtered to include events with pattern $\leq$4 and 
$\leq$12 for the pn and MOS instruments, respectively, over the energy range \hbox{0.3--10~keV}. 
High-background intervals are present in all of the observations; 
the procedure to obtain a good compromise 
between clean event files and relatively good statistics consists of using the script {\em xmmlight\_clean.csh}
\footnote{Available at http://www.sr.bham.ac.uk/xmm2/scripts.html.} 
which performs a recursive 3$\sigma$ cleaning on the lightcurves 
in the 10--15 keV energy range (binned in 100-s intervals) until the mean count rate per bin is constant. 
The goodness of this cleaning process (proven to be effective in past observations) 
has been checked on the resulting lightcurves and images, 
adopting different bin intervals, and by comparison with other cleaning methods 
(e.g., adopting the procedure reported in Baldi et al. 2002), confirming the achievement of 
a relatively good final result. The only exception is the last observation carried out in 2007 (OBS\_ID=0505050301), 
which still shows the presence of significant residual flares after the 
cleaning process, hampering a good analysis of the data. 
For this reason, the data of this observation will not be considered further. 
The final ``cleaned'' exposure times are reported in Table~\ref{obs_log}. 

To extract \xray\ spectra, we used variable source-extraction aperture radii to maximize the signal-to-noise (S/N) ratio 
over the entire 0.3--10~keV energy range and obtain good counting statistics for moderate-quality spectral analysis. 
Radii vary from 27 to 45 arcsec for the pn, and from 24 to 30 arcsec for the two MOS cameras. 
The background was taken from circular regions, in the same chip (to avoid significant spatial variations 
across the detector) and close to \pg\ (without being contaminated by the target itself) 
with radii of 90\arcsec. The adoption of multiple background regions 
did not provide significantly different results. 
%
\begin{figure*}
\includegraphics[angle=-90,width=0.48\textwidth]{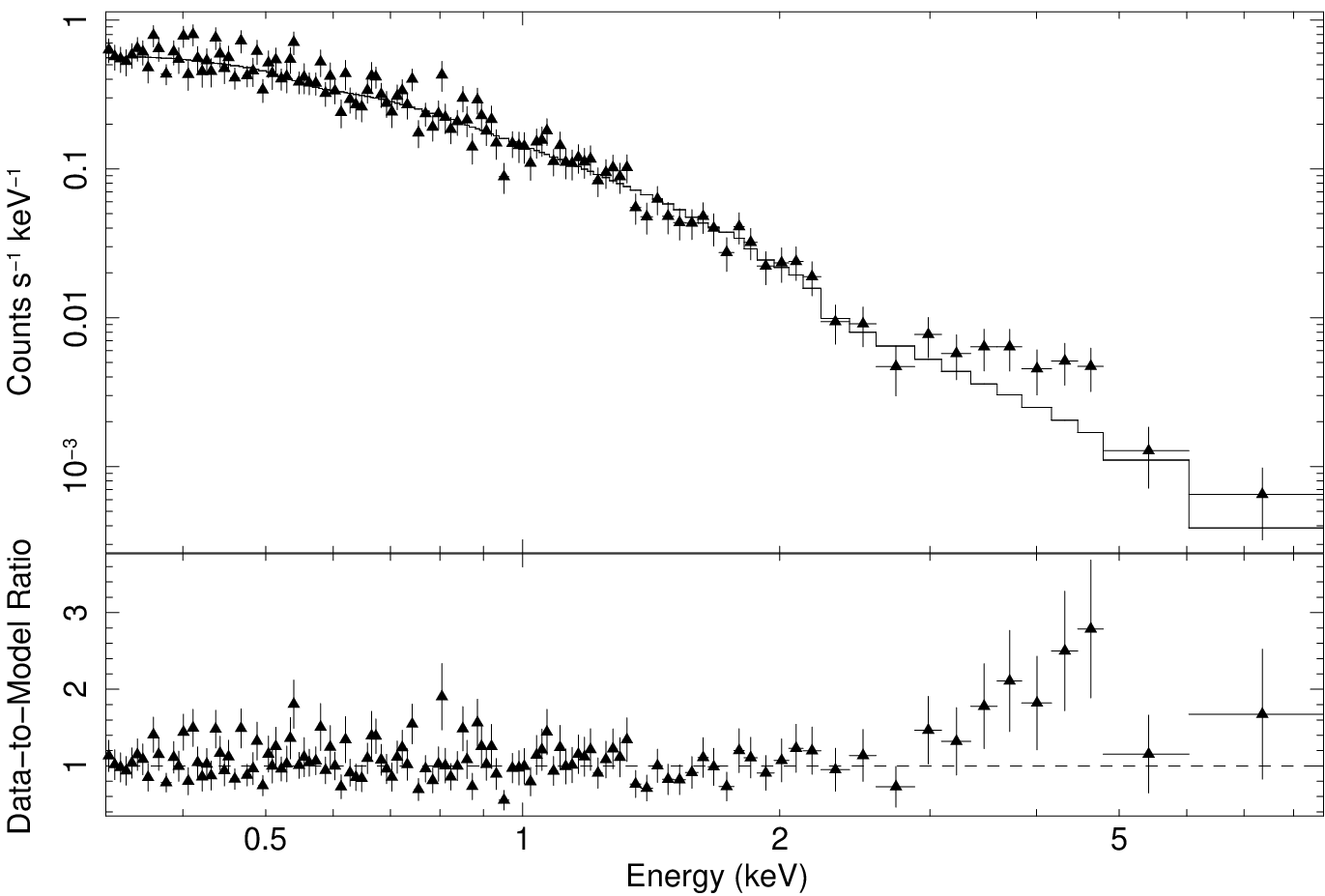}
\hfill
\includegraphics[angle=-90,width=0.44\textwidth]{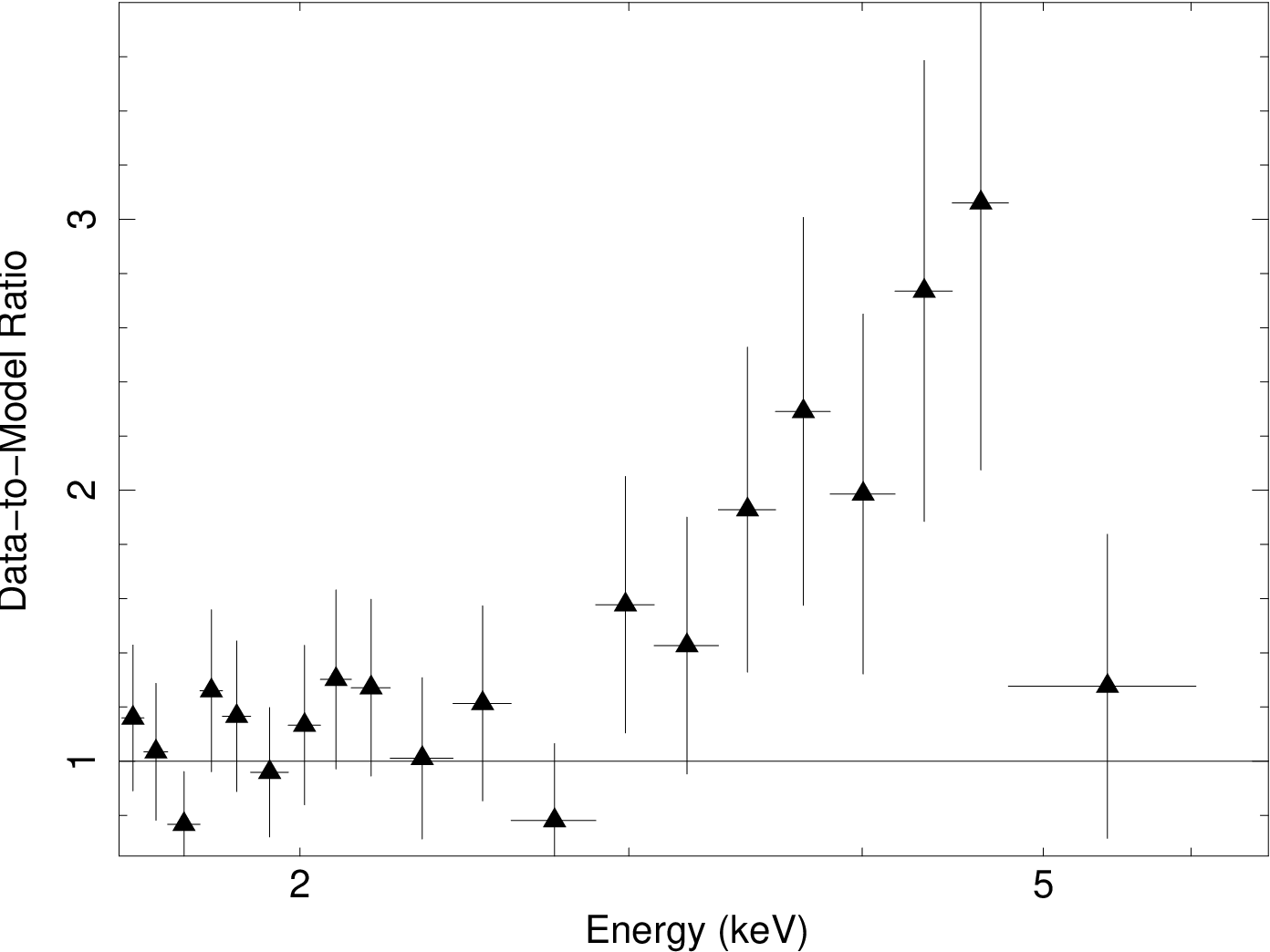}
\caption{
{\it (Left panel)} Fitting to the archival pn data (OBS\_ID=0153220401) over the 
$\approx$~0.3--10~keV energy range using a power-law model and Galactic absorption. 
The data-to-model ratios are shown in the bottom panel in units of $\sigma$. 
{\it (Right panel)} Close-up of the deviations (in units of $\sigma$) in the $\approx$~1.5--7~keV band 
of the observed EPIC pn data adopting a power-law model fitted to the (0.3--3~keV) $+$ 
(6--10~keV) range and extrapolated to the whole energy interval. 
}
\label{pn_broadline}
\end{figure*}
%
The redistribution matrix files (RMFs, which include information on the detector gain and 
energy resolution), and the ancillary response files (ARFs, which include information on the 
effective area of the instrument, filter transmission and any additional energy-dependent 
efficiencies) were created with the {\sc sas} tasks {\sc rmfgen} and {\sc arfgen}, 
respectively. The resulting pn (MOS) spectra were grouped with a minimum of 20 (15) counts per bin 
using the task {\sc grppha}, in order to apply the $\chi^{2}$ statistic, and fitted with 
{\sc xspec v12.4.0} (Arnaud 1996). 

During each observation, the flux level in the soft (0.5--2~keV) and hard band (2--10~keV) 
did not show significant (i.e., above the $\approx$~20~per~cent) variations, according to a 
$\chi^2$ test.

\subsection{Optical Monitor data}
\label{om_data}
In order to obtain information on the broad-band (UV to \xray) spectral properties of \pg\ (see $\S$\ref{discussion}),  
we used the Optical Monitor (OM) products, which consist of flux densities and magnitudes in selected filters. 
While for the archival observation only data taken with the UVM2 filter 
(having an effective wavelength of 2310~\AA) are available, all of the other observations have U$-$band coverage 
(effective wavelength of 3440~\AA), and in the first observation of June 2007 (OBS\_ID=0505050201) 
also the UVM2 filter was operating.

\subsection{X-ray spectral analysis}
\label{xray_spectrum}
At first, we checked the pn, MOS1 and MOS2 data for consistency (over the  $\approx$~0.3--10~keV band) 
within each observation, assuming a power-law model, and found generally a good agreement 
in both the photon index and the overall shape of the residuals among the different EPIC instruments. 
Some residuals appear evident in the $\approx$~3--6~keV energy range, 
being particularly prominent in the pn data (likely because of its higher 
effective area compared to that of the MOS cameras). As an example, in Fig.~1 (left panel) we report the 
fitting to the pn data with a power-law model and Galactic absorption using the archival data only, 
since this observation has subsequently driven the science case of the proposed and awarded 2007 observations. 
The residuals resemble those expected in case of a broad, relativistic (because of the pronounced red wing) iron line, 
as suggested by the close-up of the deviations centered on the presumably broad iron feature shown 
in Fig.~\ref{pn_broadline} (right panel). 

Given the necessity of a proper investigation of the hard-band spectral complexities of \pg, we decided to 
re-extract the source spectra from all EPIC instruments adopting an ``optimised'' radius for each source 
extraction region (see Table~\ref{opt_region_cnts}). 
%
\begin{table}
\centering
\begin{minipage}{\textwidth}
\caption{Counts in EPIC after optimisation of the source extraction regions.}
\label{opt_region_cnts}
\begin{tabular}{ccccccc}
\hline
                & \multicolumn{3}{c}{Opt. Extr. Region (arcsec)} & \multicolumn{3}{c}{Net Source Counts} \\
OBS$\_{\rm ID}$ & pn & MOS1 & MOS2                               & pn & MOS1 & MOS2 \\
\hline
0153220401 & 16 & 26 & 21 & 2560 & {\phn}840 & {\phn}840 \\ 
0505050201 & 18 & 40 & 24 & 2830 & 1080      & {\phn}910 \\ 
0505050701 & 16 & 38 & 25 & 2590 & 1030      & {\phn}910 \\ 
\hline
\end{tabular}
\end{minipage}
\begin{minipage}{0.48\textwidth}
The source extraction regions have been chosen to optimise the signal-to-noise ratio in 
the 2--10~keV band (see $\S$\ref{xray_spectrum} for details); source counts are reported 
over the entire 0.3--7~keV energy range. 
\end{minipage}
\end{table}

%
This radius was chosen to maximize the S/N ratio over the 2--10~keV band using the task {\em eregionanalyse} 
within the {\sc sas}. 
The results from this task were checked against different choices of the background regions. 
The spectra were binned with a minimum of 20 counts per bin in all pn observations 
and 15 counts per bin in all MOS data. 
Good source signal is present in the $\approx$~0.3--7~keV band, then at higher energies the 
background starts providing a relevant contribution in the spectral fitting. 
For this reason, to provide robust spectral results, the analyses presented in the following are 
produced in this energy range, at the expense of a total of $\approx$~150 counts 
with respect to the number of counts in the \hbox{0.3--10~keV} band. 

%
\begin{figure}
\includegraphics[angle=-90,width=0.48\textwidth]{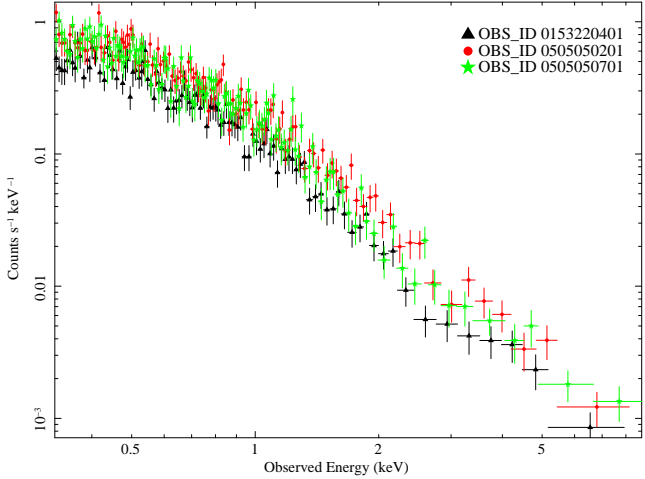}
\caption{
EPIC pn data taken from the three observations reported in Table~\ref{opt_region_cnts}, 
where the radius of the source extraction regions has been chosen such to maximize the signal-to-noise ratio 
in the 2--10~keV band (see $\S$\ref{xray_spectrum} for details).}
\label{spectra_3pn_opt}
\end{figure}
%
Despite the observed source flux variations over the rest-frame time-scale of $\approx$~3 years probed by our 
observations (the 0.5--10~keV flux was $\approx5\times10^{-13}$~\cgs\ in 2003 and increased by $\approx$~50\% 
in the observations carried out in 2007), the \xray\ spectral shape of \pg\ does not appear to vary dramatically 
(see Fig.~\ref{spectra_3pn_opt}). 
Motivated by these considerations and in order to increase the significance of our spectral results, 
we fitted all the \xray\ data (see Table~\ref{opt_region_cnts}) simultaneously, leaving 
the normalisations free to vary to account for the different calibrations (a few per cent) 
of the EPIC cameras. 
The spectral analysis presented in the following sections 
has been carried out on the summed spectra (i.e., one spectrum for three pn datasets and 
similarly for the MOS1 and MOS2 instruments) 
obtained combining the individual source (background) spectra with the tool {\em mathpha} and averaging the 
RMF and ARF matrices by the relative exposure times using the {\sc ftools} {\em addrmf} and {\em addarf}, respectively. 
This choice, accurately checked against simultaneous fitting of all the individual pn and MOS spectra, was found 
to produce reliable \xray\ spectral results and, overall, allows a more direct visualization of the spectral findings 
in case of many datasets. 

Hereafter, the quoted errors on the derived model parameters correspond to the 90~per~cent confidence level 
for one interesting parameter (i.e., $\Delta\chi^{2}=2.71$; Avni 1976), if not stated otherwise; 
all spectral fits include absorption due to the line-of-sight Galactic column density of 
\hbox{$N_{\rm H}=1.59\times10^{20}$~cm$^{-2}$} (Dickey \& Lockman 1990), and Anders \& Grevesse (1989) 
abundances are assumed.

%
\begin{table*}
\centering
\begin{minipage}{0.68\textwidth}
\caption{X-ray spectral results of the EPIC pn and MOS data.}
\label{xmm_spectra}
\begin{tabular}{cccccccc}
\hline
Model        &        $\Gamma$        &           E            &        EW              & $\xi$   &       N$_{\rm H}$      &         Cov.~Fract     & $\chi^{2}$/dof \\
(1)          &           (2)          &          (3)           &        (4)             &  (5)    &           (6)          &            (7)         &      (8)        \\
\hline
\sf (a)      & 2.70$\pm{0.03}$        &                        &                        &         &                        &                        & 495.5/442 \\ 
%
%
\sf (b)      & 2.81$^{+0.07}_{-0.04}$ & 4.09 unc.              & 2.35$^{+5.61}_{-1.23}$ &         &                        &                        & 445.0/439 \\ 
\sf (c)      & 2.75$\pm{0.03}$        & 6.71$^{+0.21}_{-0.53}$ & 3.09$\pm{0.84}$        &         &                        &                        & 458.9/440 \\ 
\sf (d)      & 2.72$^{+0.02}_{-0.03}$ & 6.33$\pm{0.11}$        & 1.12$^{+0.20}_{-0.60}$ &         &                        &                        & 476.4/440 \\
\sf (e)      & 2.35$^{+0.01}_{-0.02}$ &                        &                        & $>6760$ &                        &                        & 460.8/441 \\
\sf (f)      & 2.78$\pm{0.04}$        &                        &                        &         & 2.12$^{+1.91}_{-0.89}$ & 0.53$^{+0.12}_{-0.10}$ & 445.1/440 \\
\hline
\end{tabular}
\end{minipage}
\begin{minipage}{0.68\textwidth}
The relative normalisation of the MOS cameras with respect to the pn 
is 1.03--1.04 in all the spectral fittings presented here. 
(1) Reference for the model adopted in the spectral fitting (see below and $\S$\ref{xray_spectrum} for details); 
(2) Photon index over the 0.3--7~keV bandpass; 
(3) Rest-frame energy of the line (keV); 
(4) Rest-frame equivalent width of the line (keV); 
(5) Ionization parameter $\xi = \frac{4\pi F_{\rm tot}}{n_{\rm H}}$ (erg~cm~s$^{-1}$), 
where $F_{\rm tot}$ is the total illuminating flux and $n_{\rm H}$ is the hydrogen number density (part~cm$^{-3}$) of the illuminated slab; 
(6) Rest-frame column density (in units of 10$^{23}$~cm$^{-2}$); 
(7) Covering fraction of the absorbing matter; 
(8) Goodness of the fit in terms of $\chi^{2}$/(number of degrees of freedom, d.o.f.). \\
Adopted models: 
{\sf (a)} powerlaw; 
{\sf (b)} powerlaw and broad Gaussian iron line; 
{\sf (c)} powerlaw and relativistic ({\sc laor}) iron line; 
{\sf (d)} powerlaw and relativistic ({\sc diskline}) iron line; 
{\sf (e)} ionized reflection ({\sc reflion} model), convolved with a relativistic blurring kernel from the Laor (1991) code 
({\sc kdblur}); 
{\sf (f)} powerlaw plus partial covering absorption.
\end{minipage}
\end{table*}

\begin{figure}
\includegraphics[angle=-90,width=80mm]{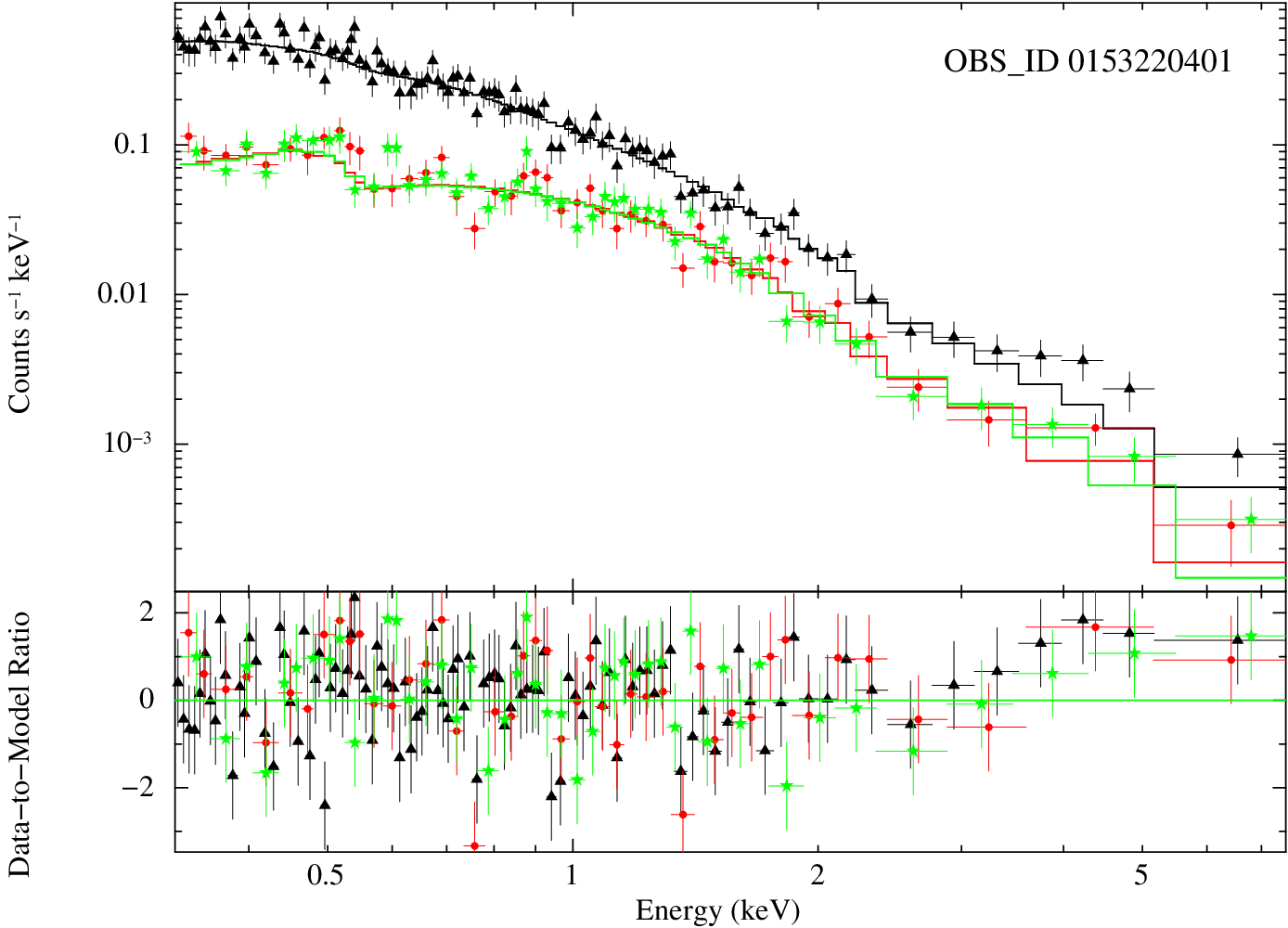}
\includegraphics[angle=-90,width=80mm]{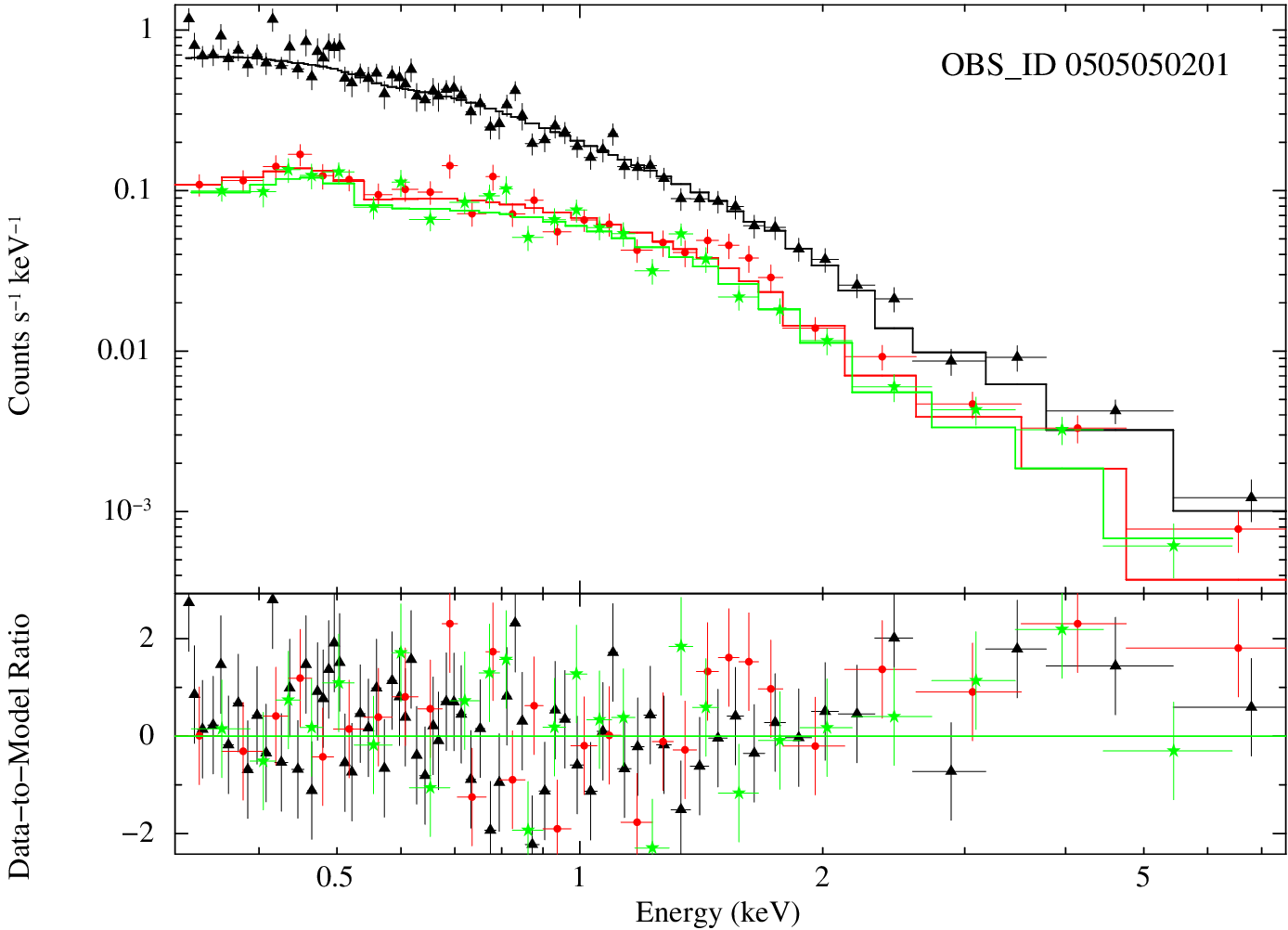}
\includegraphics[angle=-90,width=80mm]{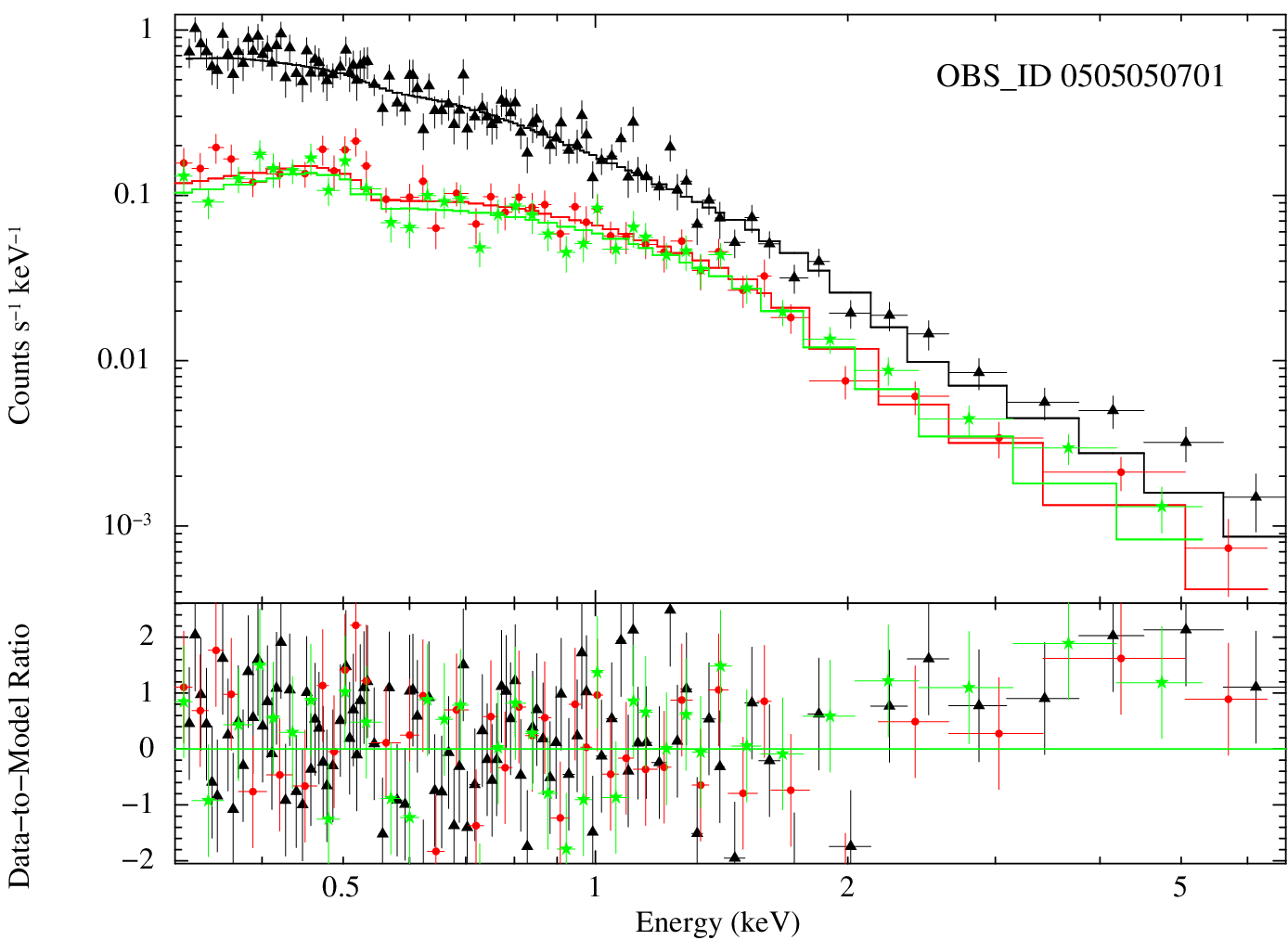}
\caption{Fitting to the pn (triangles), MOS1 (filled circles) and MOS2 (stars) spectral data of the three 
observations over the $\approx$~0.3--7~keV energy range using a power-law model and Galactic absorption. 
The data-to-model ratios are shown in the bottom panels in units of $\sigma$.}
\label{epic_po_only_3obs}
\end{figure}

\begin{figure}
\includegraphics[angle=-90,width=80mm]{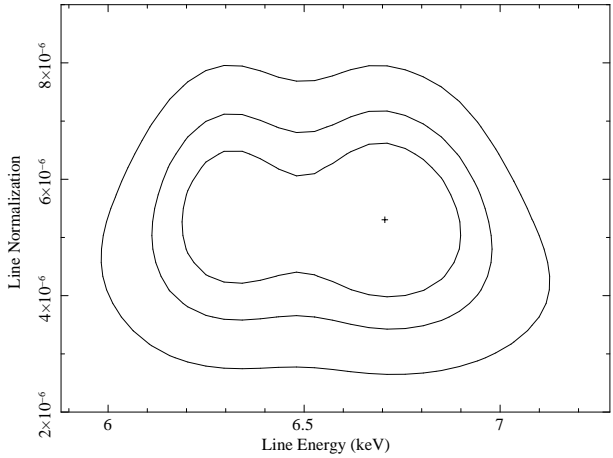}
\caption{68, 90 and 99~per~cent confidence contours showing the rest-frame energy 
vs. photon flux of the relativistic iron emission line ({\sc laor} model in {\sc xspec}).}
\label{pn_laor}
\end{figure}

%
Similarly to the situation illustrated in Fig.~\ref{pn_broadline}, a single power-law model is not 
able to reproduce the data of all the EPIC data, leaving strong residuals in the hard band 
(see Table~\ref{xmm_spectra}, model {\sf (a)} and Fig.~\ref{epic_po_only_3obs}).
%
The inclusion of a broad line, either modelled with a Gaussian 
(model {\sc zgauss} in {\sc xspec}; model {\sf (b)} in Table~\ref{xmm_spectra}) 
or a relativistic component (models {\sc diskline} and {\sc laor} in the case of a black hole in the 
Schwarzschild and Kerr metric, respectively -- models {\sf (c)} and {\sf (d)} in Table~\ref{xmm_spectra}; 
Fabian et al. 1989; Laor 1991) improves the fit 
by $\Delta\chi^{2}\approx$19--50 for 2--3 additional degrees of freedom 
(the line energy, width and normalisation in the {\sc zgauss} model, and line energy 
and normalisation in the relativistic models, where the inner radius of the accretion disc 
was fixed to the minimum radius allowed by the two metrics 
and the emissivity index was left to its default value in the {\sc laor} model). 
Although the highest improvement, in terms of $\chi^{2}$, is obtained using a Gaussian iron line, 
the width of this line ($\sigma=2.98^{+4.55}_{-1.14}$~keV), 
coupled with its centroid energy (unconstrained), 
suggests that a more accurate spectral parameterization (e.g., a relativistic feature) 
is needed to properly reproduce the data. 
Using the {\sc laor} model {\sf (c)}, the improvement in the spectral fitting corresponds to the $>$99.999~per~cent 
(according to the F-test) with respect to the single power-law model; 
since the iron line normalisation was allowed to be negative in the spectral fitting, 
the F-test can still be considered a reliable method to derive the statistical significance of 
the line (see Protassov et al. 2002). 
The best-fitting rest-frame energy, 6.7$^{+0.2}_{-0.5}$~keV in the case of a {\sc laor} line, 
is consistent, within the errors, with that obtained in case of a {\sc diskline}, with the former model 
being preferred on a statistical basis (Table~\ref{xmm_spectra}) and from visual inspection of the data-to-model ratios. 
The line energy is consistent with neutral iron up to Fe\ {\sc xxv}. 
The equivalent width (EW) of the relativistic line (EW=3.1$\pm{0.8}$~keV in the source rest frame)  
is much larger than that measured in case of detection of broad features 
by \asca\ and \xmm\ for a sample of nearby AGN (\hbox{EW$\approx$~230--250~eV}; Nandra et al. 1997; 
Guainazzi, Bianchi \& Dov{\v c}iak 2006); 
we defer a more exhaustive discussion on possible causes of this strong feature 
in $\S$\ref{discussion}. 
The rest-frame line energy vs. normalisation in the {\sc laor} model, shown 
in Fig.~\ref{pn_laor}, is likely suggestive of a more complex modeling for \pg. 
%

The power-law photon index does not vary significantly with the inclusion of a broad feature, being 
$\Gamma\approx2.7-2.8$, in agreement with that derived from \asca\ data. 
Although warm absorbers and soft excesses are often detected in PG quasars 
(e.g., Reynolds 1997; Porquet et al. 2004; Piconcelli et al. 2005), we must note that 
there is no strong indication that 
the main features of warm absorbers (i.e., the \ovii\ and \oviii\ absorption edges and the 
unresolved transition array, UTA; e.g., Piconcelli et al. 2005) are present in the EPIC spectra; 
furthermore, the source redshift ($z$=0.400) is such to plausibly move a significant fraction of any 
additional soft component below the energy interval where the EPIC instruments are mostly sensitive. 

\begin{figure}
\includegraphics[angle=-90,width=80mm]{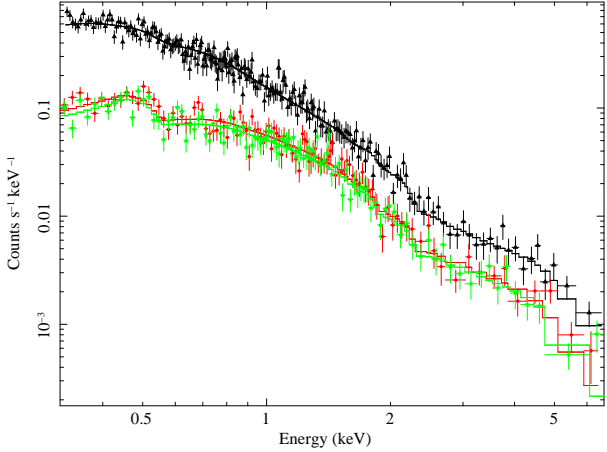}
\caption{pn (triangles), MOS1 (filled circles) and MOS2 (stars) spectral data of the three 
observations discussed in $\S$\ref{xray_spectrum} fitted with a ionized reflection model 
convolved with a relativistic blurring kernel from the Laor code 
(model {\sf (e)} in Table~\ref{xmm_spectra}).}
\label{spectrum_all_kdblur_reflion}
\end{figure}

Given the recent claims that NLS1 and NLQ \xray\ spectra could be reflection-dominated 
(e.g., Fabian et al. 2002, but see also Done 2007) and the convincing results 
obtained for some AGN of this class, where the whole \xray\ continuum emission 
and iron K$_{\alpha}$ line are well explained in terms of 
reflection (e.g., Fabian et al. 2004; Gallo et al. 2004a and references therein), 
we tried to fit all the data using a ionized reflection model 
({\sc reflion}\footnote{Available at 
http://heasarc.gsfc.nasa.gov/docs/xanadu/xspec/models/reflion.html 
as external table.} into {\sc xspec}; 
Ross \& Fabian 2005), which incorporates both line emission with Compton broadening and reflection continuum, 
convolved with a relativistic blurring kernel from the Laor (1991) code ({\sc kdblur} into {\sc xspec}; 
model {\sf (e)} in Table~\ref{xmm_spectra}). 
This spectral parameterization is well suited to represent the light-bending model, where 
the presence of a reflection-dominated spectrum is interpreted as due to strong light-bending effects 
at work close to the black hole (Miniutti \& Fabian 2004; see $\S$\ref{discussion}). 
The iron abundance was fixed to solar, and the other spectral parameters frozen to their default values. 
The quality of the fit ($\chi^{2}/d.o.f.=460.8/441$) is similar to that obtained in the case 
of a relativistic iron line plus a steep \xray\ continuum, the only difference being a flatter photon index 
($\Gamma$=2.35$^{+0.01}_{-0.02}$) and, likely, a more physical description of the \xray\ emission of \pg; 
the spectrum is shown in Fig.~\ref{spectrum_all_kdblur_reflion}
%
%

Unfortunately, the quality of the data does not allow us to constrain the ionization parameter 
($\xi = \frac{4\pi F_{\rm tot}}{n_{\rm H}}>6760$~erg~cm~s$^{-1}$, where $F_{\rm tot}$ is the total illuminating flux 
and $n_{\rm H}$ is the hydrogen number density) and to distinguish, on the basis of the $\chi^{2}$ and 
the inner radius of the accretion disc (once left free to vary), between a maximally spinning black hole 
and a non-spinning Schwarzschild black hole. 
In these highly ionised conditions, the disc acts like a perfect reflector (see Ross \& Fabian 2005), 
whose main properties are a power-law similar to the incident one, a modest Compton hump, and a 
``smeared'' iron line and edge, due to the effects of Comptonization. For this reason, 
it is difficult to place constraints on the relative strength of the
reflection continuum with respect to the incident powerlaw, which is broadly constrained 
in the range 1.5--6.0 for solar abundances. 
Although the optical spectrum indicates iron over-abundance (see $\S$2.3 of Aoki et al. 2005), we 
note that, from an \xray\ perspective, a slightly better statistical result 
(improvement by $\Delta\chi^{2}\approx$~9 for one additional degree of freedom with respect to the model 
with iron abundance fixed to solar) is achieved leaving the iron 
abundance free to vary in the spectral fitting (and the Fe abundance relative to the solar 
value is 0.64$^{+0.26}_{-0.25}$). 

We also note that the iron edge at $\approx$~7.1~keV or sharp drop in the continuum flux 
observed in some NLS1s (e.g., Boller 2004; Gallo et al. 2004b) and interpreted in terms of 
reflection cannot be investigated by the current data, because at these high energies the source and 
background signals become approximately comparable (see $\S$\ref{xray_spectrum}). 

Finally, we tried to model the spectrum of \pg\ using a partial-covering model 
(model {\sf (f)}) instead of a reflection component. 
This spectral fitting provides a statistically better result ($\chi^{2}/d.o.f.=445.1/440$), 
a steep photon index \hbox{($\Gamma=2.78\pm{0.04}$)} and 
a covering fraction (i.e., the fraction of the nuclear source covered by the absorber) of 0.53$^{+0.12}_{-0.10}$. 
The derived rest-frame column density (\hbox{N$_{\rm H}=2.12^{+1.91}_{-0.89}\times10^{23}$~cm$^{-2}$}) 
appears difficult to reconcile with the apparent lack of extinction in the optical/UV spectrum of \pg\ 
(e.g., Baskin \& Laor 2005) but could be explained 
assuming a dust-to-gas ratio largely below the Galactic value (e.g., Maiolino et al. 2001). 
In this model, where the absorber must be close to the \xray\ emitting region, possibly within the sublimation 
radius (thus explaining the low extinction), we would expect to observe significant variations in the parameters 
of the partial covering model over the three observations used in the spectral fitting. 
Although the spectral constraints from the individual observations are relatively poor, we find that 
in the archival observation and in the two 2007 observations the column density is 
\hbox{N$_{\rm H}=3.2^{+5.5}_{-2.0}\times10^{23}$~cm$^{-2}$} and \hbox{N$_{\rm H}=1.2^{+1.3}_{-0.5}\times10^{23}$~cm$^{-2}$}, 
and the covering fraction  0.59$\pm{0.27}$ and 0.53$^{+0.11}_{-0.14}$, respectively. 
Therefore, within the statistical uncertainties, it seems that no significant variations in the 
absorber occur over a rest-frame time-scale of $\approx$~3~years, which might cast some doubts on the 
physical reliability of this model. 
On the other hand, in the framework of the light-bending model, a reduced 
variability of the reflection component with respect to the continuum can be 
explained. If the continuum variability is not entirely intrinsic but is 
mainly due to changes in the location of the primary source of hard X-rays, 
light-bending effects close to the central massive black hole predict little 
variability of the spectral components reprocessed by the accretion disc 
(Miniutti \& Fabian 2004).

In all the models mentioned above, the inclusion of a narrow ($\sigma$=10~eV) 
iron K$_{\alpha}$ line at 6.4~keV (rest-frame EW$<$450~eV) is not required by the data. 

%
%

\subsection{Long-term X-ray flux variability of \pg}
\label{longterm_flux}
In the attempt to evaluate the long-term behaviour of \pg, we analysed the 
\asca\ on-axis observation of this quasar (see George et al. 2000), using the 
products available at the {\sc tartarus} 
database\footnote{See http://tartarus.gsfc.nasa.gov/.}), 
and an off-axis observation with \chandra, coupled with the three \xmm\ observations presented 
in the previous section. 
From \asca\ to the archival \xmm\ observation, over a time-scale of $\approx$6~years 
($\approx$4.3 years in the source rest frame), the \hbox{2--10~keV} source flux 
has changed by a factor of $\approx$3.5 (from $\approx$5.1--5.3$\times10^{-13}$~\cgs\ in 
the \asca\ observation to $\approx$1.7$\times10^{-13}$~\cgs\ in the first \xmm\ observation, which corresponds 
to a rest-frame \hbox{2--10~keV} luminosity of $\approx$1.1$\times10^{44}$~\lum), 
with no evidence for \xray\ spectral variability. 
Unfortunately, the 2002 \chandra\ observation does not provide enough counts ($\approx$~220) 
for a detailed \xray\ spectral fitting, and the spectral analysis is limited in the energy range 
$\approx$~0.5--4 keV. Within the uncertainties, the constraints on the photon index obtained 
by \chandra\ are consistent with our \xmm\ results, and the extrapolated \hbox{2--10~keV} flux 
is intermediate between those measured by \asca\ and \xmm\ (archival observation).  
In 2007, the source 2--10~keV flux increased to $\approx$2.7$\times10^{-13}$~\cgs\ (OBS$\_{\rm ID}$=0505050201) 
and by a similar factor in the 0.5--2~keV band, and became $\approx$2.4$\times10^{-13}$~\cgs\ in the last analysed 
\xmm\ observation (OBS$\_{\rm ID}$=0505050701).

\section{Discussion}
\label{discussion}
From a broad-band perspective, the combined analysis of EPIC and OM data of \pg\ allows us 
a simultaneous study of its SED, parameterized by \aox, the slope of 
the hypothetical powerlaw connecting the rest-frame wavelengths of 
2500~\AA\ and 2~keV and defined as 
\begin{equation}
\alpha_{\rm ox}=\frac{\log(f_{\rm 2~keV}/f_{2500~\mbox{\rm \scriptsize\AA}})}{\log(\nu_{\rm 2~keV}/\nu_{2500~\mbox{\rm \scriptsize\AA}})}
\end{equation}
where $f_{\rm 2~keV}$ and $f_{2500~\mbox{\scriptsize \rm \AA}}$ are the 
flux densities at rest-frame 2~keV and 2500~\AA, respectively. 
In this regard, the derived \aox\ values ($-$1.64 for the archival 
observation, $-$1.45 and $-$1.47 for the 2007 
observations analysed in $\S$\ref{xray_spectrum}\footnote{For the 2007 observations, 
in the \aox\ calculation we used the U$-$band flux density, 
since this filter at $z$=0.40 approximately samples the rest-frame wavelength of 2500~\AA, while for the 
archival observation, the flux density obtained from the UVM2 filter was extrapolated assuming the UV slope 
reported in Matsumoto et al. (2006); our result is consistent with Matsumoto et al. findings.}) 
are fully consistent with those expected on the basis of the known anti-correlation between this spectral index and 
the 2500~\AA\ luminosity (using the most up-to-date parameterization reported in Just et al. 2007). 
Therefore, at least in the UV-to-X-ray energy range, 
the broad-band properties of \pg\ do not appear to be unusual. 

From an \xray\ point-of-view, the spectrum of \pg\ appears characterized by a steep photon index and a strongly skewed, 
relativistic iron line, both accounted for by a ionized reflection model. 
There are indications that the former spectral result may be related to the high accretion rate of 
the source (Aoki et al. 2005; Baskin \& Laor 2005), as actually found in most of the NLS1s and NLQs 
(e.g., Boroson 2002). 
As a further support to such indication, Shemmer et al. (2006), using a sample of 
30 quasars up to redshift $\approx$~2, have shown that the photon index depends primarily on the 
source accretion rate; this correlation is more robust than that obtained for 
$\Gamma$ vs. FWHM(H$_\beta$), i.e., sources with steeper \xray\ spectra (both in the soft 
and hard X-rays) have typically narrower H$_\beta$ emission lines (e.g., Laor et al. 1994, 1997; 
Brandt et al. 1997), as actually observed also in \pg. 

As a non-secondary effect/indication of the high accretion rate of this source, 
there is evidence for a strong blueshift and asymmetric profile in the \oiii\ 5007\AA\ 
and \civ\ 1459\AA\ emission lines (Aoki et al. 2005; Baskin \& Laor 2005); 
a high-ionization, optically thin wind, propagating outward the broad-line region 
and caused by the radiation pressure of the quasar represents a viable explanation (Marziani et al. 2003b), 
as well as a scenario whereby the narrow-line region clouds are entrained in a decelerating wind 
(Komossa et al. 2008). 
This result finds support from the dependence of the amount of blueshift vs. the source optical 
luminosity in a sample of quasars with similar properties to those of \pg\ (Aoki et al. 2005). 

The presence of a relativistic iron line places this quasar among the small number of 
luminous AGN where such feature has been detected so far. 
In addition to those Seyfert-like AGN that ``historically'' (i.e., since \asca\ and \rxte\ observations) 
were known to doubtlessly show relativistic iron lines (e.g., MCG$-$6$-$30$-$15, Tanaka et al. 1995; 
NGC~3516, Nandra et al. 1999; IRAS~18325$-$5926, Iwasawa et al. 1996; MCG$-$5$-$23$-$16, Weaver, Krolik \& Pier 1998), 
over the last few years \xmm\ and \chandra\ 
good-quality spectral data have allowed detailed investigations of the presence of 
relativistic features in a sizable number of AGN (e.g., Porquet \& Reeves 2003; 
Miniutti \& Fabian 2006; Porquet 2006; Piconcelli et al. 2006 and references therein; 
Longinotti et al. 2007; Miniutti et al. 2006; Krumpe et al. 2007; Schartel et al. 2007), 
also at moderately high redshifts (e.g., Comastri, Brusa \& Civano 2004; Chartas et al. 2007), 
and have enabled relevant statistical analyses (Guainazzi et al. 2006; Nandra et al. 2006; Inoue, Terashima \& Ho 2007). 
From these studies, $\approx$~40~per~cent of the \xmm\ AGN with more than 
10000 counts in the 2--10~keV band have broad iron features (Guainazzi et al. 2006), although the 
situation may be more complex and foresee the presence of complex absorption (e.g., 
Nandra et al. 2007). 

Although further data may be required to provide better constraints on the ionization of the 
continuum and, possibly, on the line parameters, 
it is interesting to note that this is the first time that such line is detected in \pg: 
the \asca\ observation showed only marginal evidence for the presence of a broad iron line, 
probably due to the higher background level of the \asca\ observation and lower effective area 
of its instruments. 
We also note that the line was not detected in the analysis of the archival \xmm\ 
observation carried out by Brocksopp et al. (2006) and Matsumoto et al. (2006), 
with an upper limit to a narrow iron K$_{\alpha}$ line being 800~eV according to the latter authors. 
It seems plausible that the data reduction procedure and cleaning performed by the other authors 
are different from ours. This possibility is confirmed by the much larger statistical uncertainties that 
Brocksopp et al. (2006) quote in their spectral fittings than those presented in this work 
($\S$\ref{xray_spectrum}), probably caused by a less effective subtraction of the flaring-background 
intervals (as apparent from their Fig.~1) and by our adopted strategy to enhance the S/N ratio 
in the hard band via the source extraction radius optimisation ($\S$\ref{xray_spectrum}). 
Moreover, the availability of multiple \xmm\ observations of \pg\ showing a similar spectral behaviour 
provides further support to our results. 

A possible cause of concern is the large EW of the iron line (EW=3.1$\pm{0.8}$~keV in the source rest frame), 
much larger than typically observed in local Seyfert galaxies and quasars and among the largest EWs ever found 
in AGN (see, for comparison, Fig.~2 of Guainazzi et al. 2006). 
Plausible explanations include 
(1) significant iron overabundance, 
(2) reflection-dominated spectrum which, even with solar Fe abundance, 
yields to a self-consistent line EW with respect to its own reflection continuum, 
(3) matter partially covering the \xray\ source. 
Although iron over-abundance is present in the optical spectrum of \pg\ (Aoki et al. 2005), 
the fact that the iron line intensity grows logarithmically with the iron abundance 
(Matt, Fabian \& Reynolds 1997) would require an extremely high abundance to account for the 
observed iron K$_{\alpha}$ line EW. Therefore, this hypothesis appears unlikely. 

The second hypothesis, the possibility that the \xray\ spectrum is reflection-dominated, 
appears the most plausible, although more accurate spectral data are required to provide a more 
robust and convincing support to it. According to recent literature, 
the presence of a reflection-dominated spectrum can be associated with strong light-bending 
effects at work close to the black hole (Miniutti \& Fabian 2004). 
In this model, strong light bending is expected if the primary source is located close 
to the central black hole and illuminates the inner regions of the accretion disc. 
In this situation, almost all of the radiation emitted by the source is bent onto the disc 
rather than escaping to the observer. 
In general, the relative fraction of reflected vs. observed power-law flux depends 
on the height of the source above the disc (e.g., see Miniutti et al. 2003; 
Miniutti, Fabian \& Miller 2004; Crummy et al. 2005; Ponti et al. 2006). 
Given the good fit to the data obtained using a ionized reflection model convolved with a 
relativistic blurring kernel from the Laor (1991) code, 
the source of \xray\ photons should be located very close to the black hole during the \xmm\ observations 
presented here. This reflection model predicts broad and strong iron lines (also see Dabrowski \& Lasenby 2001 for 
details) when the source is in a relatively low \xray\ flux regime, 
which could be the case for \pg\ in the \xmm\ observations (see $\S$\ref{longterm_flux}); furthermore, it has provided 
a plausible explanation for the spectral properties of some NLS1s showing reflection-dominated 
spectra and, possibly, strong iron lines (e.g., Miniutti \& Fabian 2004; Porquet 2006), for the 
luminous NLQ PHL~1092 (Gallo et al. 2004a) and, recently, for the soft excess of many PG quasars 
(Crummy et al. 2006). 
As recently pointed out by Merloni et al. (2006), reflection-dominated spectra could also be 
produced in a disc consisting of a inhomogeneously heated plasma (hot phase) 
pervaded by small dense clumps (cold phase), which intercept most of the photons coming 
from the hot phase. Therefore, while in this model the reflection is related to the clumpy and 
inhomogeneous nature of the inner disc, in the light-bending model it is due to general relativistic 
effects (for details, see Malzac, Merloni \& Suebsuwong 2006), although any further sign of such effects on 
the \xray\ emission is probably not visible because of the still limited statistics. 
Unfortunately, the current \xray\ data do not allow us to discriminate between these two models. 

From a statistical and physical basis, also a partial-covering model (third hypothesis) 
provides a good description of the \xray\ spectral complexities of \pg. 
Under some assumptions for the absorbing medium (see $\S$\ref{xray_spectrum}), 
it is possible to reproduce the optical properties of \pg, which shows a blue continuum 
and features related to the broad-line region, with no indications for significant extinction. 
In this model, if the absorber is in form of clouds close to the nuclear engine, 
\xray\ spectral variability is likely to take place due to orbital motions of the matter 
around the black hole and/or outflow phenomena (see $\S$\ref{introduction}). 
The lack of any appreciable variations in the properties of the absorber over the time-scale probed 
by our observations suggests that this model is less likely than the light bending model; however, 
we note that the large uncertainties in the measurements derived from the individual observations prevent us from 
obtaining a firm conclusion on this issue.  

Over the next years, with the availability of extended observations of 
NLS1s and NLQs, it would be interesting to understand whether some sort of connection 
exists between the presence of ionized reflection (related to the closeness of the source of \xray\ 
photons to the black hole in the light-bending model), the accretion rate of the AGN and winds/outflows phenomena. 

\section{Summary}
We have analysed an archival plus proprietary \xmm\ observations of \pg\ 
(probing a rest-frame time-scale of $\approx$~3 years), a narrow-line quasar at $z$=0.400. 
We found evidence for a steep continuum and a relativistic iron K$_{\alpha}$ emission line, which 
are both accounted for in the framework of either ionized reflection or partial-covering of the source. 
The large EW (3.1$\pm{0.8}$~keV in the source rest frame) of the line would be naturally explained 
in the context of ionized reflection if the source of \xray\ photons is very close to the accreting black hole, 
thus being subject to strong gravity effects; however, a partial covering model cannot be ruled out 
on the basis of the current data. 
The steep \xray\ spectrum is possibly related to the high accretion rate of \pg, as suggested 
by recent studies on quasars and, likely, by the strong outflow and asymmetric profiles measured in the \oiii\ 5007\AA\ 
and \civ\ 1549\AA\ emission lines, which make of \pg\ one of the most intriguing quasars in the PG sample 
and a valuable scientific case for future investigations. 

\section*{Acknowledgments}
CV, EP and SB thank for support the Italian Space Agency (contracts ASI--INAF I/023/05/0 and ASI I/088/06/0); 
GM acknowledges support from the UK PPARC. 
The authors thank J. Fritz, M. Meneghetti, M. Mignoli, L. Moscardini, G. Ponti and O. Shemmer 
for useful discussions, and the anonymous referee for his/her useful comments and suggestions which improved the 
quality of the paper.

\end{document}